\documentclass[conference]{IEEEtran}
\IEEEoverridecommandlockouts
\usepackage{cite}
\usepackage{amsmath,amssymb,amsfonts}
\usepackage{graphicx}
\usepackage{textcomp}
\usepackage{xcolor}
\usepackage{algorithm}
\usepackage{hyperref}
\usepackage{algpseudocode}
\usepackage[table]{xcolor}
\usepackage{booktabs}
\def\BibTeX{{\rm B\kern-.05em{\sc i\kern-.025em b}\kern-.08em
    T\kern-.1667em\lower.7ex\hbox{E}\kern-.125emX}}
\algnewcommand{\ParFor}[1]{\textbf{for all} #1 \textbf{do}}
\algnewcommand{\EndParFor}{\algorithmicend\ \textbf{for}}

\begin{document}

\title{APEX: A Dual-Sparsity Accelerator for Precise and Efficient 
SNN Inference}

\author{
\IEEEauthorblockN{
Devgokul Bawa Venkatesh, Sreeram Radhakrishnan, Rajshekhar Rakshit, Gopalakrishnan Srinivasan
}
\IEEEauthorblockA{
Department of Computer Science and Engineering\\
Indian Institute of Technology Madras\\
Chennai, India\\
\texttt{\{ns26z056,ee23b075\}@smail.iitm.ac.in, \{cs24s031,sgopal\}@cse.iitm.ac.in}
}
}

\maketitle

\begin{abstract}
Spiking Neural Networks (SNNs) have emerged as an energy-efficient alternative to Artificial Neural Networks (ANNs), leveraging sparse accumulate operations in the place of power-hungry multiply-and-accumulate operations. ANN-SNN conversion is a widely adopted approach to realize deep SNNs with accuracy comparable to that of ANNs. The Quantization-Clip-Floor-Shift (QCFS) activation minimizes conversion error, yet requires a large number of inference timesteps to match the source ANN accuracy on real-world vision datasets. PASCAL addresses this by proposing the Precise ANN-SNN Conversion Integrate-and-Fire (PASC-IF) neuron, which guarantees mathematical equivalence between the converted SNN and the source ANN, thereby achieving ANN-equivalent accuracy at significantly reduced timesteps. Despite this algorithmic advancement, the hardware implications of deploying the PASC-IF neuron remain unexplored. In this work, we present APEX, a dual-sparsity SNN inference accelerator that integrates the PASC-IF neuron into the LoAS hardware framework. The three-stage PASC-IF datapath is realized as a fully combinational circuit with no additional latency cost. APEX exploits dual sparsity in both input spikes and weights through a fully temporal-parallel dataflow, enabling efficient sparse computation and reduced memory traffic. Across all evaluated models, the PASC-IF neuron on average achieves up to 3\% higher accuracy than the standard IF neuron, with a power overhead of only 1.3\%–5.4\%, an area overhead of 2.1\%–2.7\%, and 40\% energy reduction for best accuracy configurations.
\end{abstract}

\section{Introduction}

Spiking Neural Networks (SNNs) have attracted significant attention as an energy-efficient alternative to Artificial Neural Networks (ANNs) for edge inference~\cite{cheng2024snn}. By encoding information as binary spikes and replacing multiply-accumulate (MAC) operations with low-cost accumulate (AC) operations, SNNs provide substantial energy savings on dedicated hardware~\cite{yin2024loas},\cite{spinalflow2020},\cite{stellar2024}. ANN-SNN conversion is the dominant approach for obtaining high-accuracy deep SNNs without the prohibitive cost of direct training using backpropagation-through-time~\cite{wu2018spatio}. QCFS activation~\cite{bu2023optimal} minimizes conversion error by replacing ReLU with a quantized activation function during ANN training. However, the converted SNN still requires up to ${\sim}1024$ timesteps to match source ANN accuracy on complex datasets such as ImageNet. PASCAL~\cite{ramesh2025pascal} addresses this by introducing the PASC-IF neuron, which replaces the standard integrate-fire (IF) neuron with a three-stage spike-count mechanism that is mathematically proven to reproduce the QCFS activation output exactly, guaranteeing ANN-level accuracy at any timestep budget.

Despite this algorithmic advance, no existing SNN accelerator supports the PASC-IF neuron. Its three-stage execution, inhibitory spike paths, and soft-reset mechanism are incompatible with conventional IF neuron hardware. 
A naive sequential implementation would reintroduce temporal dependencies that undermine the efficiency of modern parallel SNN dataflows. 
In this work, we present APEX, a dual-sparsity SNN inference accelerator that integrates the PASC-IF neuron. APEX is built upon LoAS~\cite{yin2024loas}, which adopts a fully temporal-parallel (FTP) inner-product dataflow to exploit dual sparsity in both spikes and weights. The key observation is that the FTP dataflow fully resolves the post-synaptic input tensor before the neuron unit is invoked, eliminating the temporal dependency that would otherwise prevent parallel execution of the PASC stages. This allows all three stages to be unrolled as combinational circuits, each completing in a single clock cycle. In addition, APEX supports mixed-precision execution through an evenly split INT4/INT8 processing-element array. Our contributions are as follows:
\begin{itemize}
    \item On average, APEX achieves 40\% energy reduction over LoAS, for the best accuracy configurations. Across all evaluated models, APEX consistently achieves competitive or higher accuracy than LoAS at reduced timestep and energy budgets, with a power overhead of 1.3\%--5.4\% and an area overhead of 2.1\%--2.7\%.
    \item We present the hardware implementation of the PASC-IF neuron as a fully combinational 3-cycle datapath, which is integrated into the LoAS accelerator with no changes to the surrounding architecture.
    \item APEX exploits dual sparsity in both spikes and weights via an FTP dataflow, and supports mixed-precision INT4/ INT8 execution, reducing energy cost while maintaining high utilization.
\end{itemize}

\section{Background and Motivation}

\subsection{Spiking Neural Networks and the IF Neuron}

\begin{figure}[t]
    \centering
    \includegraphics[width=\linewidth]{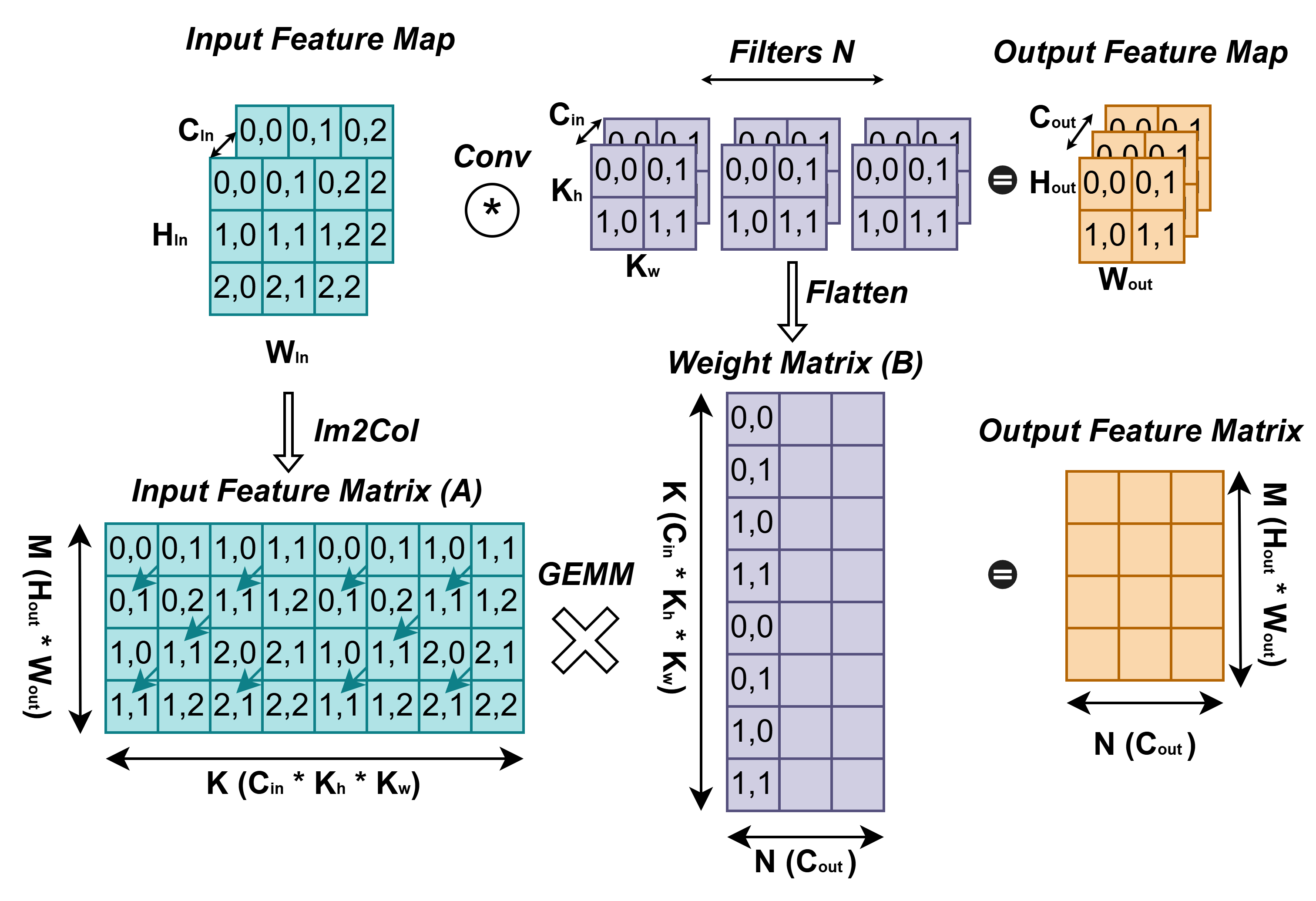}
    \caption{Im2Col transformation reduces both convolutional and fully-connected layers to a unified GEMM operation. The input feature map is reshaped into matrix $A \in \{0,1\}^{M \times K}$, where $M = H_{out} \times W_{out}$ and $K = C_{in} \times k_h \times k_w$, and the filters are flattened into weight matrix $B \in \mathbb{R}^{K \times N}$, where $N = C_{out}$.}
    \label{fig:im2col}
\end{figure}

Unlike traditional ANNs, which process inputs using dense multiply-accumulate (MAC) operations, SNNs encode information as discrete binary $\{0,1\}$ spikes, enabling sparse, event-driven computation~\cite{cheng2024snn}. Among spiking neuron models, the Integrate-and-Fire (IF) neuron offers the best tradeoff between biological plausibility and computational efficiency. As shown in Fig.~\ref{fig:im2col}, both fully-connected and convolutional layers reduce to the same matrix multiplication form via an im2col transformation: the input feature map is reshaped into a matrix $A \in \{0,1\}^{M \times K \times T}$, where $M = H_{out} \times W_{out}$ is the number of output spatial locations, $K = C_{in} \times k_h \times k_w$ aggregates the input channels and kernel spatial extent ($k_h$, $k_w$ denote the kernel height and width), and $T$ is the number of discrete timesteps. The filters are similarly flattened into a weight matrix $B \in \mathbb{R}^{K \times N}$, where $N = C_{out}$ is the number of output channels. The IF operation then proceeds in two steps.

\textbf{Synaptic accumulation.} At each SNN inference timestep $t_i \in \{0,\ldots,T{-}1\}$, the post-synaptic input $O \in \mathbb{R}^{M \times N \times T}$ is computed as:
\begin{equation}
    O_{m,n}[t_i] = \sum_{k=0}^{K} A_{m,k}[t_i]\, B_{k,n}.
\end{equation}
Since $A[t_i]$ is binary and typically sparse, this reduces to a sparse-matrix sparse-matrix (spMspM) multiplication, where the expensive MAC units are replaced with low-cost accumulate (AC) operations.

\textbf{Firing and reset.} A spiking neuron maintains a membrane potential $U \in \mathbb{R}^{M \times N}$, which accumulates the synaptic inputs over time. At timestep $t_i$, the integrated membrane state $X \in \mathbb{R}^{M \times N \times T}$ is specified by
\begin{equation}
    X_{m,n}[t_i] = O_{m,n}[t_i] + U_{m,n}[t_{i-1}],
\end{equation}
where $U_{m,n}[t_{i-1}]$ carries the neuron's accumulated history from the previous timestep. An output spike is emitted when $X$ exceeds a fixed firing threshold $v_{th} \in \mathbb{R}$:
\begin{equation}
    C_{m,n}[t_i] =
    \begin{cases}
        1, & \text{if } X_{m,n}[t_i] > v_{th}, \\
        0, & \text{otherwise,}
    \end{cases}
\end{equation}
where $C \in \{0,1\}^{M \times N \times T}$ is the output spike tensor. The membrane potential is then hard reset to zero whenever a spike is emitted ($C_{m,n}[t_i] = 1$), which can be formulated as
\begin{equation}
    U_{m,n}[t_i] = X_{m,n}[t_i]\,\bigl(1 - C_{m,n}[t_i]\bigr).
\end{equation}
This temporal dependency, where an output spike at $t_i$ depends on $U[t_{i-1}]$, is the foundational challenge for parallel hardware execution, which we address in Section~\ref{sec:ftp}.

\subsection{SNN Training Methodologies}
\label{sec:training}

SNN training methodologies fall into two broad categories: direct training and ANN-SNN conversion. Direct training~\cite{wu2018spatio,deng2022temporalefficienttrainingspiking} via backpropagation-through-time (BPTT) yields competitive accuracy on complex datasets such as ImageNet~\cite{deng2022temporalefficienttrainingspiking}, with inference latencies typically below a few tens of timesteps. However, since SNNs necessitate multiple sequential forward passes over $T$ timesteps, BPTT accumulates error gradients across all timesteps, multiplying both memory and compute requirements by a factor of $T$ relative to an equivalent ANN~\cite{wu2018spatio}. This makes direct training computationally intensive for deep networks.

ANN-SNN conversion eliminates this cost by transferring pretrained ANN weights directly into an equivalent SNN, replacing each ReLU activation with an IF spiking neuron whose average rate of firing approximates the original ReLU output. Although conversion algorithms achieve accuracy comparable to the source ANN, they require inference latencies spanning hundreds to thousands of timesteps~\cite{sengupta2019going}.

\textbf{QCFS Activation~\cite{bu2023optimal}.} Bu~et~al. addressed this by replacing ReLU with Quantization-Clip-Floor-Shift (QCFS) activation during ANN training:
\begin{equation}
    \hat{h}(z^l) = \lambda^l \,\mathrm{clip}\!\left(
        \frac{1}{L}\left\lfloor \frac{z^l L}{\lambda^l} + 
        \varphi \right\rfloor, 0, 1\right),
\end{equation}
where $z^l$ is the input to the activation layer, $L$ is the quantization step, $\lambda^l$ is the trainable activation threshold, and $\varphi = \frac{1}{2}$ is a shift term that minimizes the expected conversion error. At the end of ANN training, $\lambda^l = \theta_n$, where $\theta_n$ is the firing threshold used during SNN inference. While QCFS formally minimizes the conversion error, in practice, SNNs trained with QCFS still require up to ${\sim}1024$ timesteps to match source ANN accuracy on complex datasets such as ImageNet, despite being trained with $L = 4$ or $8$ quantization steps~\cite{bu2023optimal}.

\textbf{PASCAL and the PASC Neuron~\cite{ramesh2025pascal}.} PASCAL addresses this limitation by converting a QCFS-trained ANN into an SNN that is \emph{mathematically equivalent} to the source ANN, guaranteeing ANN-level accuracy at any timestep budget. This is achieved through the Precise ANN-SNN Conversion (PASC) neuron, which replaces the standard IF neuron with a three-stage spike generation mechanism. Given the quantization step $L_{n-1}$ of the preceding layer and $L_n$ of the current layer, with layerwise threshold $\theta^* = \frac{\theta_n}{L_n}$ and initial membrane potential $\mathrm{mem}(0) = \frac{\theta^*}{2}$, the PASC neuron operates as follows:

\begin{enumerate}
    \item \textbf{Stage 1 --- Excitatory accumulation ($L_{n-1}$ timesteps).} The neuron integrates the incoming spike train using soft reset: upon firing, the membrane potential is decremented by $\theta^*$ rather than reset to zero, preserving residual potential. A spike counter $s$ records the cumulative spike count scaled by $\frac{\theta_n}{L_n}$.

    \item \textbf{Stage 2 --- Inhibitory correction ($\max(L_{n-1}, L_n) - 1$ timesteps).} The membrane potential of the neuron is re-initialized to the value at the end of Stage 1. Both excitatory and inhibitory spikes are enabled: if $\mathrm{mem}(t) \geq \theta^*$, an excitatory neuronal spike increments $s$; if $\mathrm{mem}(t) < 0$, an inhibitory spike decrements $s$ while the membrane potential is incremented by $\theta^*$. This stage is critical for correcting over- or under-accumulation errors that arise when $L_{n-1} \neq L_n$.

    \item \textbf{Stage 3 --- Output spike generation ($L_n$ timesteps).} The membrane potential is reset to the scaled spike count $s$ obtained after Stage 2. Standard IF firing is applied for $L_n$ timesteps to generate the final output spike train $s^l$, which is passed to the subsequent layer.
\end{enumerate}

PASCAL formally proves that $\sum_{t=1}^{L_n} s^l(t) = \hat{h}(z^l)$ for every neuron, thus guaranteeing that the sum of output spikes exactly reproduces the QCFS activation output~\cite{ramesh2025pascal}. Converting a QCFS-trained ANN to use the PASC-IF neuron instead of the standard IF neuron consistently yields higher accuracy at lower timestep budgets across all model architectures and weight precisions, with full results presented in Section~\ref{sec:results}. However, the three-stage execution, per-neuron spike counter, inhibitory spike paths, and soft reset mechanism of the PASC neuron are not natively supported by existing SNN hardware accelerators, which implement only the standard integrate-fire neuron. This is the central motivation for APEX.

\section{APEX Architecture}
\label{sec:apex}

\subsection{Overview}
\label{sec:overview}

\begin{figure}[t]
\centering
\includegraphics[width=\columnwidth]{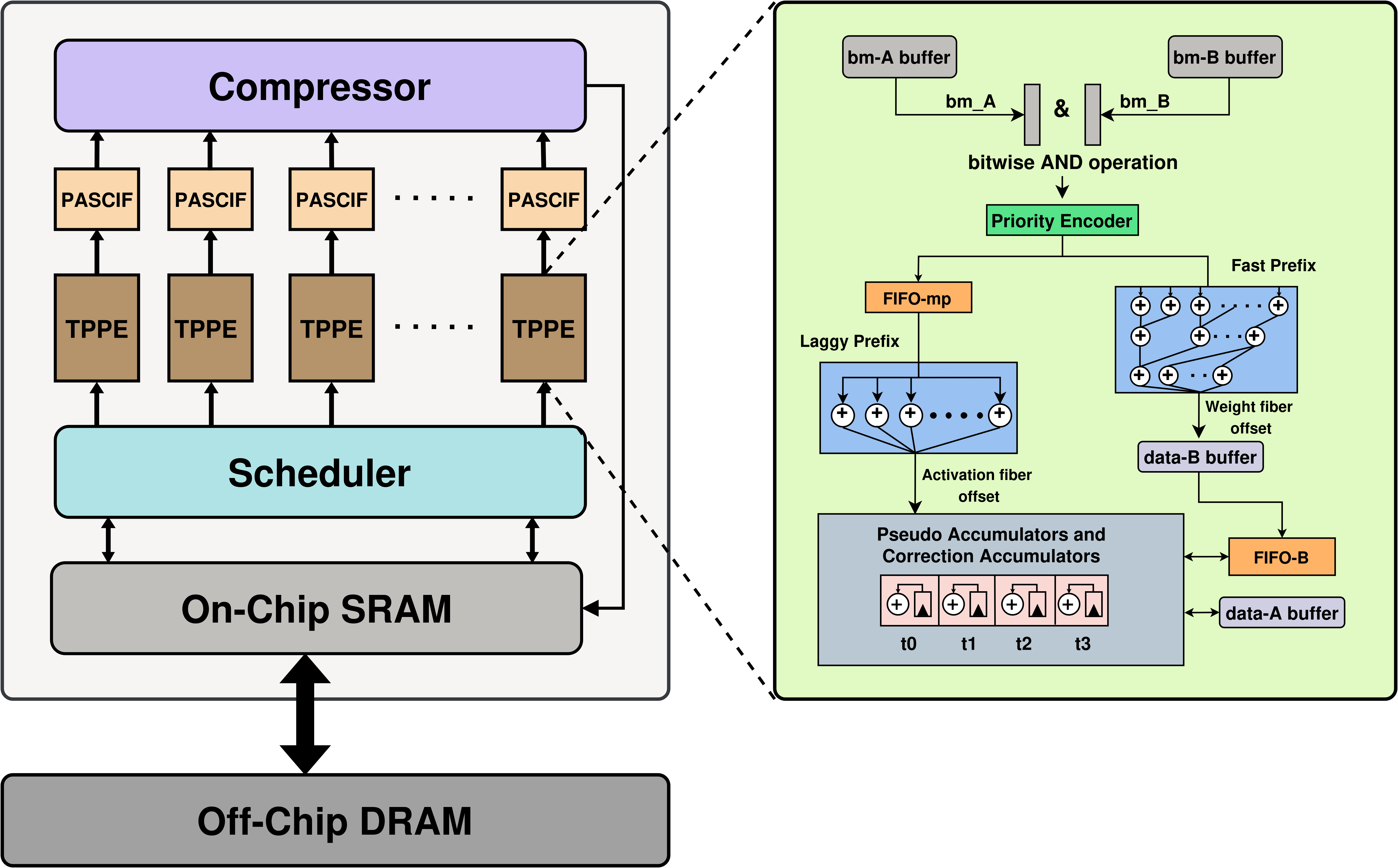}
\caption{Architecture of the APEX accelerator and TPPE microarchitecture, inspired by LoAS \cite{yin2024loas}. FIFO-mp stores matched positions from the priority encoder for later use in laggy prefix-sum computation, which takes 8 cycles while positions are generated by the priority encoder every cycle. FIFO-B holds weights fetched from the data-B buffer using offsets from the fast prefix circuit, for later use by the correction accumulators. The fast prefix is an adder tree that computes prefix sums for weight fiber access in a single cycle.}
\label{fig:APEX_ARCH}
\end{figure}

APEX is built upon the LoAS~\cite{yin2024loas} SNN accelerator architecture, which adopts a Fully Temporal Parallel (FTP) Inner-Product dataflow to exploit dual sparsity in both input spikes and weights. As illustrated in Fig.~\ref{fig:APEX_ARCH}, the core components of the architecture are the Temporal Parallel Processing Elements (TPPEs), which perform sparse matrix multiplication, and the Parallel Integrate-and-Fire (P-IF) unit, which applies neuron dynamics to generate output spikes for all timesteps in parallel. The scheduler distributes SNN workload across TPPE units, while the compressor encodes output spikes into a Compressed Sparse Fiber (CSF) representation and writes them back to the on-chip global cache.

The sole but critical modification introduced by APEX is the replacement of the standard integrate-fire neuron in LoAS with the PASC-IF neuron~\cite{ramesh2025pascal}. As established in Section~\ref{sec:training}, the standard IF neuron limits inference accuracy at low timestep budgets due to conversion errors that accumulate across layers. The PASC-IF neuron eliminates this key limitation through its three-stage spike generation mechanism, guaranteeing mathematical equivalence between the converted SNN and the source ANN. All other architectural components, namely, the FTP dataflow, CSF compression, mixed-precision TPPEs, and inner-join execution, are retained from LoAS without modification. The following subsections describe each component in detail.

\subsection{FTP Dataflow and Bit Compression}
\label{sec:ftp}

\begin{figure}[t]
\centering
\includegraphics[width=0.9\columnwidth]{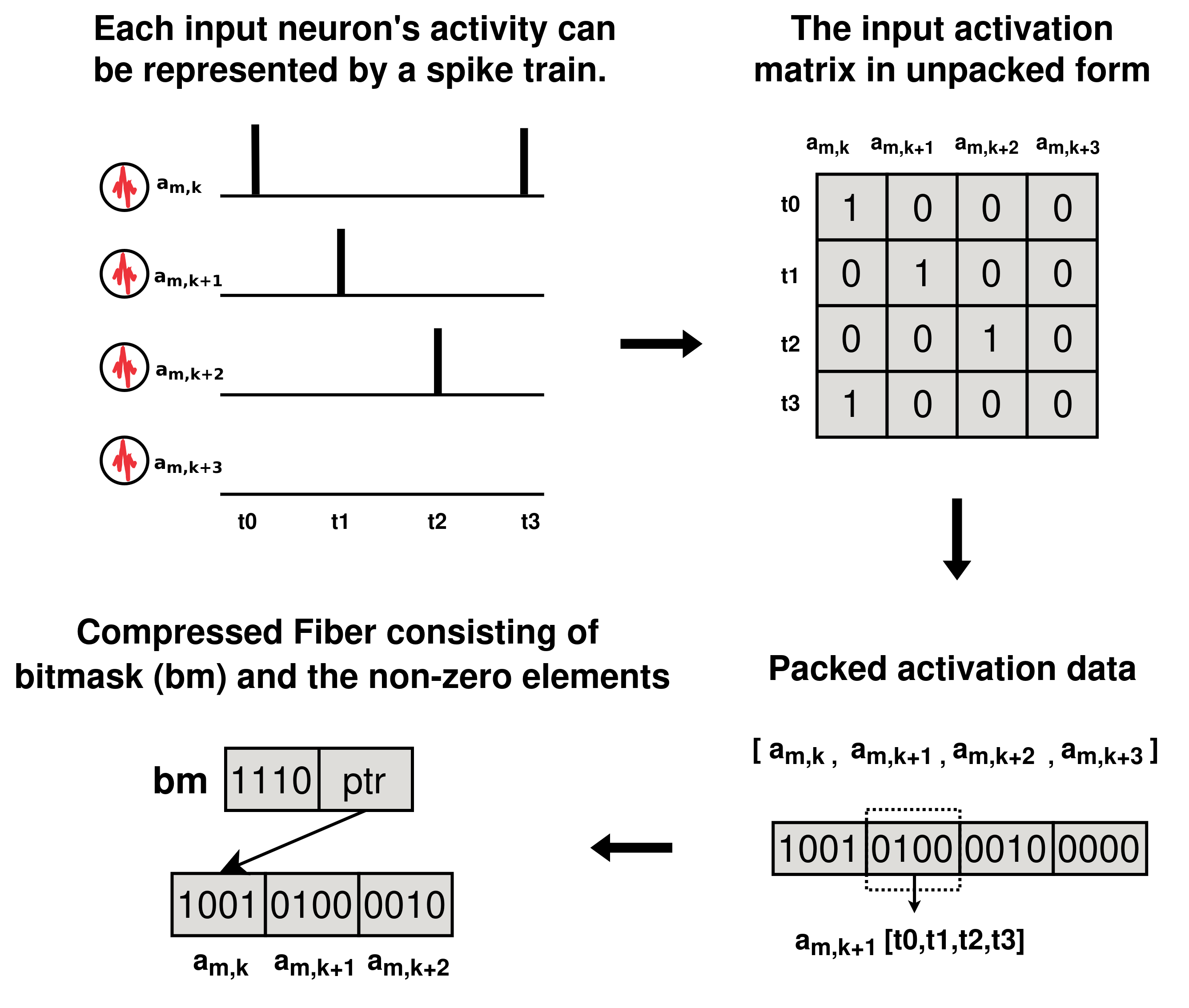}
\caption{FTP-friendly input spike compression used in APEX; \texttt{bm} refers to the bitmask and \texttt{ptr} refers to the pointer to the start address of the compressed non-zero data stored contiguously in memory.}
\label{fig:CSF}
\end{figure}

APEX adopts the Fully Temporal Parallel (or FTP) Inner-Product dataflow proposed by LoAS~\cite{yin2024loas}, wherein the temporal dimension $t$ is mapped to the innermost loop. This eliminates redundant memory accesses and avoids the accumulation of unnecessary partial sums, as in the case of
Inner-Product \textbf{(IP)}~\cite{gondimalla2019sparten}, Outer-Product \textbf{(OP)}~\cite{deng2021gospa}, and Gustavson's \textbf{(GUST)}~\cite{zhang2021gamma} dataflow, thus improving both computational and memory efficiency. 
To maximize spike compression and maintain contiguous memory access of non-zero spikes across timesteps, which effectively minimizes data movement across different levels of the memory hierarchy, APEX inherits the FTP-friendly Compressed Sparse Fiber (CSF) representation from LoAS~\cite{yin2024loas}. As shown in Fig.~\ref{fig:CSF}, the spike activity of each neuron across $T$ timesteps is encoded as a $T$-bit vector, where each bit indicates the presence or absence of a spike at the corresponding timestep. Neurons that do not generate spikes across all $T$ timesteps, referred to as \emph{silent neurons}, do not contribute to the output activation. Including such neurons in computation incurs unnecessary latency and memory traffic overhead. Hence, the compressed fiber retains only non-silent neurons, effectively reducing memory traffic and improving computational efficiency.

\subsection{Mixed Precision Enabled TPPEs}
\label{sec:tppe}

The compute engine of APEX consists of two primary components: the Temporal Parallel Processing Elements (TPPEs) and the Parallel Integrate-Fire (or P-IF) unit. In the beginning of execution, the compressed weight fiber (fiber-B), including its bitmask (bm-B) and associated non-zero values, is fetched from SRAM and broadcast to the local bitmask buffers within each TPPE. In parallel, the bitmask of the input spike fiber (bm-A) is retrieved and distributed across TPPEs, with each TPPE assigned a distinct fiber along a row of the input matrix. This distribution of both input activations and weight fibers is coordinated by the scheduler.

Once both the bitmasks are available, each TPPE performs an inner-join operation between bm-A and bm-B to identify indices where both fibers contain non-zero entries. Based on this matching, only the relevant non-zero elements from the input fiber are retrieved from the global cache and forwarded to the pseudo-accumulator, where accumulation is performed efficiently without processing zero entries. Each TPPE thereby computes the full contribution corresponding to a given output neuron across all timesteps. Upon completing accumulation, the result is forwarded to the P-IF unit, which applies spiking neuron dynamics to generate output spikes across all timesteps in parallel.

APEX supports mixed-precision execution at the filter level. The weight matrix is organized such that each column corresponds to a filter, enabling independent quantization of filters to different precisions (e.g., INT4 or INT8). To support this efficiently, the TPPE array is evenly split into INT4 and INT8 PEs. The Scheduler assigns each filter to a TPPE operating at the corresponding precision, allowing multiple precisions to be processed concurrently without incurring the overhead of fully reconfigurable units. This design enables high PE utilization while improving energy efficiency.

\subsection{Inner Join Operation}
\label{sec:innerjoin}

Each TPPE performs an inner-join operation between the bitmask of the input spike fiber (bm-A) and the bitmask of the weight fiber (bm-B) to identify positions where both fibers contain non-zero entries. A logical AND is first applied to the two bitmasks to produce the AND-result, which identifies matched positions where both the input spike and the weight are non-zero. Only the corresponding non-zero elements are then forwarded to the pseudo-accumulator for the accumulate operation, completely skipping all zero-valued entries.

A naive implementation of this inner-join requires two fast prefix-sum circuits, one for each bitmask, for computing the memory offsets of matched non-zero values. However, a single fast prefix-sum circuit alone accounts for over 45\% of system-level power, as reported in LoAS~\cite{yin2024loas}. Our APEX inherits the FTP-friendly inner-join design presented in LoAS~\cite{yin2024loas}, which replaces one of the two fast prefix-sum circuits with a \emph{laggy} prefix-sum circuit, exploiting the asymmetry between weight and spike inputs in SNNs. Since input spikes are binary $\{0,1\}$ values, the accelerator need not wait for the exact spike value before beginning to accumulate the matched weight --- it opportunistically presumes that the matched non-zero value in fiber-A fires at all timesteps and begins accumulating fiber-B immediately via the fast prefix-sum circuit. The laggy prefix-sum circuit subsequently verifies the actual spike values and issues corrections via dedicated correction accumulators where needed. This design nearly halves the cost of the prefix-sum circuitry with negligible throughput penalty, as the latency of fetching fiber-B overlaps with the correction process.

\subsection{PASC-IF Neuron Unit}
\label{sec:pascif}

The P-IF module in LoAS receives the post-synaptic input tensor $O \in \mathbb{R}^{M \times N \times T}$ from the TPPEs and applies the spiking dynamics across all $T$ timesteps in parallel. APEX replaces the IF neuron (supported by LoAS) in this unit with the PASC-IF neuron, whose three-stage spike generation mechanism is described in Section~\ref{sec:training}. 
The key hardware insight is that the FTP dataflow (Section~\ref{sec:ftp}) fully resolves post-synaptic input $O$ across all timesteps before the neuron unit is invoked, eliminating the temporal loop-carried dependency that would otherwise require sequential execution. This allows each of the three PASC stages to be implemented as a fully combinational, unrolled circuit, in which successive iterations of each stage's loop are chained directly through adder and comparator logic, completing in a single clock cycle. All $M \times N$ neurons are processed in parallel, one stage per cycle, for a total neuronal latency of 3 cycles compared to 1 cycle for the standard P-IF unit. The output spike train $C_{m,n}[1..L_n]$ produced by stage-3 is forwarded to the compressor, encoded into CSF format, and written back to the on-chip global cache for the subsequent layer.

The 2-cycle latency overhead is small in practice, as end-to-end inference latency is dominated by the TPPE accumulation pipeline. The power overhead is similarly modest, since the IF  layer contributes far fewer operations than the preceding MatMul layer~\cite{ramesh2025pascal}. Full characterization is presented in Section~\ref{sec:results}.

\begin{figure*}[t]
    \centering
    \includegraphics[width=\textwidth]{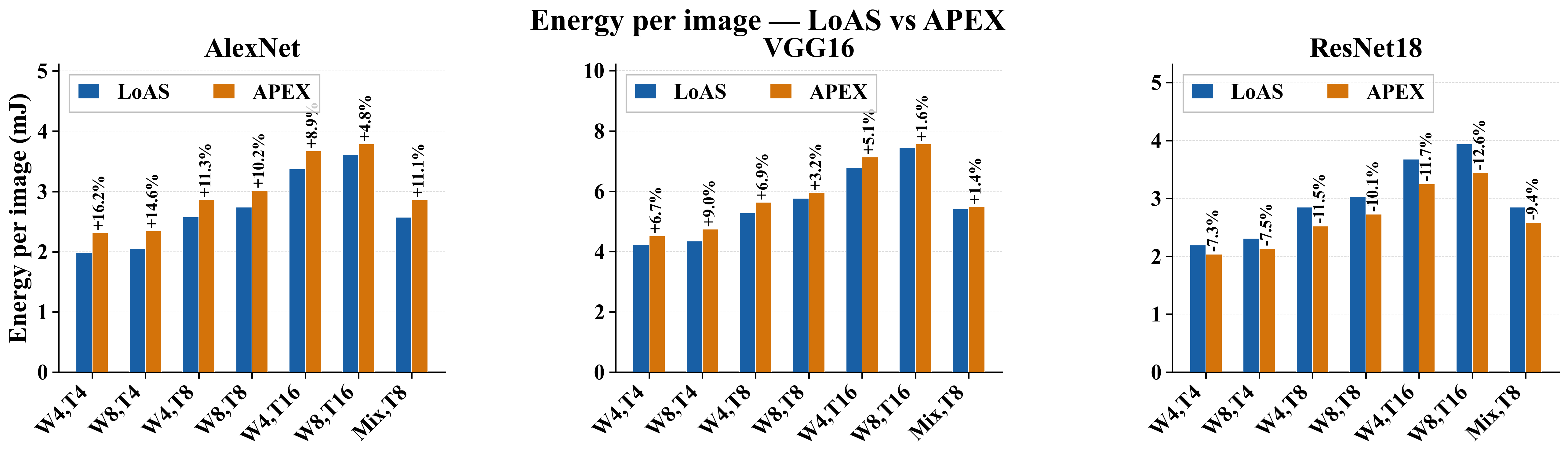}
    \caption{Energy per image (mJ) for LoAS and APEX across all models and hardware configurations. Percentage labels indicate the overhead or reduction of APEX relative to LoAS.}
    \label{fig:energy}
\end{figure*}

\begin{table}[t]
\centering
\caption{Accuracy, weight sparsity (Wt.Sp.), and activation sparsity (Act.Sp.) for QCFS and PASC-IF converted SNNs under INT8, INT4, and mixed-precision quantization on CIFAR-10.}
\label{tab:accuracy}
\setlength{\tabcolsep}{2.5pt}
\renewcommand{\arraystretch}{1.1}
\footnotesize
\begin{tabular}{llccrrrrrr}
\toprule
& & & &
\multicolumn{2}{c}{\textbf{Wt.Sp.(\%)}} &
\multicolumn{2}{c}{\textbf{Act.Sp.(\%)}} &
\multicolumn{2}{c}{\textbf{Acc.(\%)}} \\
\cmidrule(lr){5-6}\cmidrule(lr){7-8}\cmidrule(lr){9-10}
\textbf{SNN} & \textbf{Prec.} & \textbf{T} &
& \textbf{QCFS} & \textbf{PASC} &
  \textbf{QCFS} & \textbf{PASC} &
  \textbf{QCFS} & \textbf{PASC} \\
\midrule

AlexNet & INT8 & 4  && 27.54 & 1.82  & 80.97 & 79.31 & 91.56 & 93.40 \\
        & \cellcolor{yellow!20}INT8 & \cellcolor{yellow!20}8  && \cellcolor{yellow!20}27.54 & \cellcolor{yellow!20}1.82  & \cellcolor{yellow!20}77.41 & \cellcolor{yellow!20}76.09 & \cellcolor{yellow!20}\textbf{92.69} & \cellcolor{yellow!20}\textbf{93.77} \\
        & INT8 & 16 && 27.54 & 1.82  & 75.26 & 74.08 & 93.22 & 93.81 \\
        & INT4 & 4  && 62.09 & 25.83 & 80.93 & 79.01 & 87.78 & 91.52 \\
        & INT4 & 8  && 62.09 & 25.83 & 77.20 & 75.72 & 89.80 & 91.98 \\
        & INT4 & 16 && 62.09 & 25.83 & 74.98 & 73.79 & 90.30 & 92.17 \\
        & \cellcolor{green!15}Mixed & \cellcolor{green!15}8 && \cellcolor{green!15}50.67 & \cellcolor{green!15}19.01 & \cellcolor{green!15}77.39 & \cellcolor{green!15}76.04 & \cellcolor{green!15}\textbf{92.69} & \cellcolor{green!15}\textbf{93.61} \\
\midrule

VGG16   & INT8 & 4  && 14.33 & 6.64  & 75.51 & 75.00 & 93.18 & 95.39 \\
        & \cellcolor{yellow!20}INT8 & \cellcolor{yellow!20}8  && \cellcolor{yellow!20}14.33 & \cellcolor{yellow!20}6.64  & \cellcolor{yellow!20}71.40 & \cellcolor{yellow!20}71.36 & \cellcolor{yellow!20}\textbf{94.96} & \cellcolor{yellow!20}\textbf{95.61} \\
        & INT8 & 16 && 14.33 & 6.64  & 69.09 & 69.08 & 95.66 & 95.69 \\
        & INT4 & 4  && 49.33 & 43.12 & 75.71 & 75.14 & 89.79 & 94.60 \\
        & INT4 & 8  && 49.33 & 43.12 & 71.55 & 71.35 & 92.48 & 95.03 \\
        & INT4 & 16 && 49.33 & 43.12 & 69.23 & 69.09 & 93.38 & 95.13 \\
        & \cellcolor{green!15}Mixed & \cellcolor{green!15}8 && \cellcolor{green!15}34.36 & \cellcolor{green!15}24.51 & \cellcolor{green!15}71.37 & \cellcolor{green!15}71.49 & \cellcolor{green!15}\textbf{94.94} & \cellcolor{green!15}\textbf{95.41} \\
\midrule

ResNet18 & INT8 & 4  && 9.18  & 1.97  & 69.70 & 73.24 & 88.05 & 96.03 \\
         & \cellcolor{yellow!20}INT8 & \cellcolor{yellow!20}8  && \cellcolor{yellow!20}9.18  & \cellcolor{yellow!20}1.97  & \cellcolor{yellow!20}63.78 & \cellcolor{yellow!20}69.09 & \cellcolor{yellow!20}\textbf{92.20} & \cellcolor{yellow!20}\textbf{96.49} \\
         & INT8 & 16 && 9.18  & 1.97  & 60.57 & 66.58 & 94.20 & 96.65 \\
         & INT4 & 4  && 53.36 & 30.89 & 70.11 & 73.41 & 85.25 & 95.72 \\
         & INT4 & 8  && 53.36 & 30.89 & 64.25 & 69.29 & 90.59 & 96.02 \\
         & INT4 & 16 && 53.56 & 30.89 & 61.08 & 66.70 & 92.67 & 96.16 \\
         & \cellcolor{green!15}Mixed & \cellcolor{green!15}8 && \cellcolor{green!15}43.28 & \cellcolor{green!15}21.43 & \cellcolor{green!15}63.76 & \cellcolor{green!15}69.08 & \cellcolor{green!15}\textbf{92.24} & \cellcolor{green!15}\textbf{96.50} \\
\bottomrule
\end{tabular}
\end{table}

\begin{table}[t]
\centering
\caption{Accuracy, weight sparsity (Wt.Sp.), and activation sparsity (Act.Sp.) for QCFS and PASC-IF converted SNNs under INT8 quantization on the ImageNet dataset.}
\label{tab:accuracy_imagenet}
\setlength{\tabcolsep}{2.5pt}
\renewcommand{\arraystretch}{1.1}
\footnotesize
\begin{tabular}{llccrrrr}
\toprule
& & & &
\textbf{Wt.Sp.(\%)} &
\textbf{Act.Sp.(\%)} &
\textbf{Acc.(\%)} \\
\textbf{SNN} & \textbf{Method} & \textbf{T} & & & & \\
\midrule
ResNet34 (INT8) & \cellcolor{green!15}PASC & \cellcolor{green!15}8  && \cellcolor{green!15}2.72 & \cellcolor{green!15}69.09 & \cellcolor{green!15}\textbf{74.30} \\
                & QCFS    & 16 && 2.72 & 42.60 & 58.79          \\
                & \cellcolor{yellow!15}QCFS & \cellcolor{yellow!15}32 && \cellcolor{yellow!15}2.72 & \cellcolor{yellow!15}41.22 & \cellcolor{yellow!15}\textbf{69.22} \\
\bottomrule
\end{tabular}
\end{table}

\section{Results}
\label{sec:results}

\subsection{Software Configuration}

We evaluate APEX using AlexNet~\cite{AlexNet}, VGG16~\cite{VGG16}, and ResNet18~\cite{ResNet} on CIFAR-10, across timestep configurations $T \in \{4, 8, 16\}$, and ResNet34 on ImageNet. Each model is converted to an SNN using two spiking neuron configurations: the standard QCFS-converted IF neuron~\cite{bu2023optimal} and the PASC-IF neuron~\cite{ramesh2025pascal}, the latter guaranteeing mathematical equivalence to the source ANN. All models are evaluated under INT8, INT4, and mixed-precision weight quantization, applied post-training on a per-layer basis. Mixed-precision assignment follows a layerwise sensitivity analysis, and achieves accuracy comparable to uniform INT8 precision while improving energy efficiency through increased weight sparsity, as shown in Table~\ref{tab:accuracy}.

The average spike sparsity reported in Table~\ref{tab:accuracy} is computed as the fraction of silent neurons, i.e., those producing no spike across the entire inference window, over the total number of neurons in the activation matrix, as described in Section~\ref{sec:ftp}.

\subsection{Hardware Configuration}

APEX is implemented in RTL and synthesized using Synopsys Design Compiler at 400\,MHz on 40\,nm CMOS technology. 
APEX comprises 16 TPPEs, evenly split between INT4 and INT8 processing elements to support mixed-precision execution. Each TPPE contains one inner-join unit, five accumulators (one 12-bit pseudo-accumulator and four 10-bit correction accumulators), and a PASC-IF neuron unit replacing the standard IF module of LoAS. A 256\,KB double-buffered global cache with 16 banks provides sufficient on-chip data reuse across all evaluated workloads. A 128\,GB/s HBM module serves as off-chip memory. We evaluate APEX across timestep configurations $T \in \{4, 8, 16\}$ and weight precisions INT4 and INT8.


We compare APEX against LoAS~\cite{yin2024loas} as our primary hardware baseline, evaluating power, area, and energy efficiency. Since APEX retains all architectural components of LoAS and introduces only the PASC-IF neuron unit and the mixed-precision TPPE split, any overhead relative to LoAS is directly attributable to these two modifications. Results are presented in Section~\ref{sec:hw_results}.

\subsection{Hardware Results}
\label{sec:hw_results}

\begin{figure}[t]
    \centering
    \includegraphics[width=\linewidth]{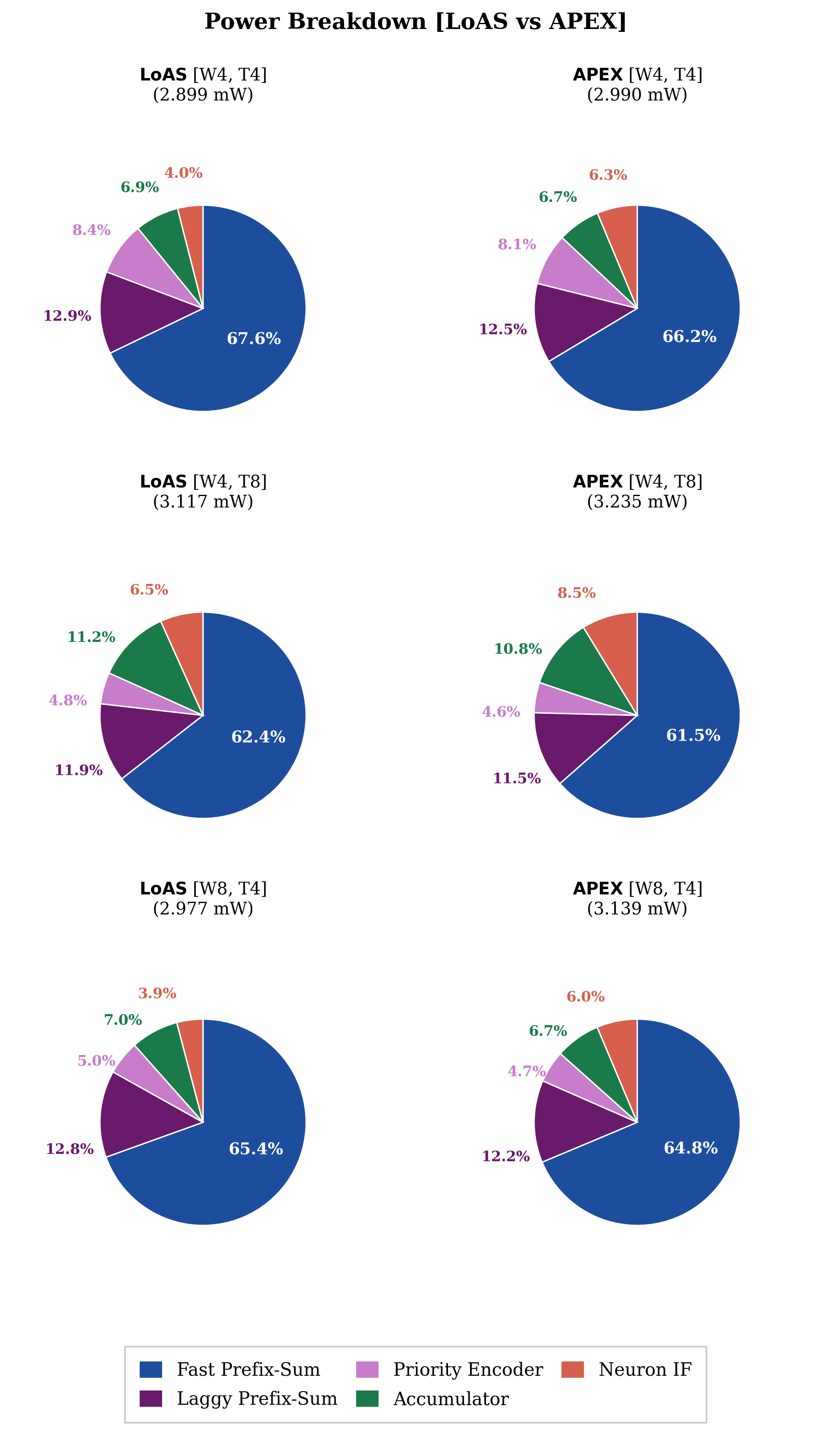}
     \caption{Power breakdown of LoAS and APEX across W4T4, W4T8, and W8T4 configurations. The fast prefix-sum circuit dominates in both designs. The PASC-IF neuron unit draws marginally more power than the standard IF neuron, reflecting the additional combinational logic of the three-stage datapath.}
    \label{fig:pie}
\end{figure}

\subsubsection{Power and Area Overhead}

\begin{figure}[t]
    \centering
    \includegraphics[width=\linewidth]{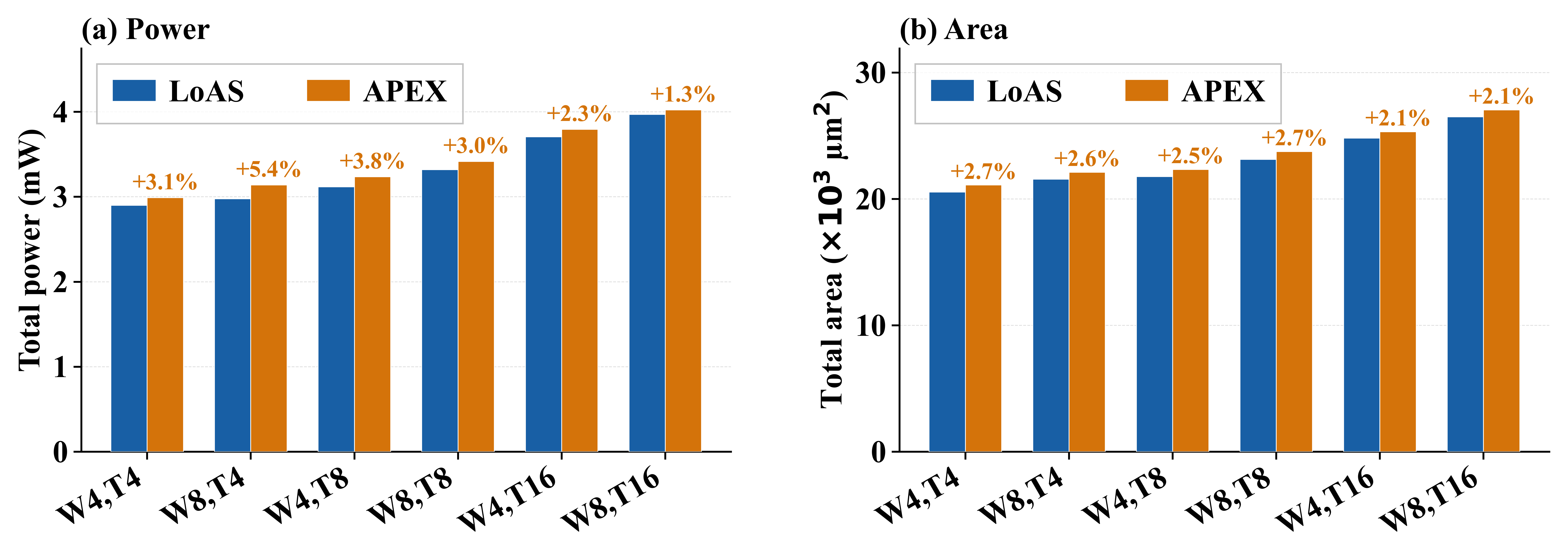}
    \caption{Total power (mW) and area ($\times 10^3\,\mu\text{m}^2$) of LoAS and APEX across all weight precision and timestep configurations. Percentage labels indicate the overhead of APEX relative to LoAS.}
    \label{fig:power_area}
\end{figure}

Fig.~\ref{fig:power_area} compares the total power and area of APEX against LoAS across all six hardware configurations, where weight precision $W \in \{4, 8\}$ bits and timesteps $T \in \{4, 8, 16\}$). APEX incurs a power overhead of 1.3\%--5.4\% and an area overhead of 2.1\%--2.7\% relative to LoAS, confirming that ANN-equivalent inference accuracy is achieved at minimal hardware cost.

\subsubsection{Power Breakdown}

Fig.~\ref{fig:pie} shows the power breakdown of LoAS and APEX for W4,T4 configuration. The fast prefix-sum circuit dominates power consumption in both the designs, accounting for 67.60\% in LoAS and 66.20\% in APEX. The PASC-IF neuron unit contributes 6.3\% of the total power in APEX, compared to 4.0\% for the standard IF neuron in LoAS, reflecting the additional combinational logic of the three-stage datapath. The proportions of all the other components remain largely unchanged between the two designs.

\subsubsection{Energy Efficiency}

\begin{figure*}[t]
    \centering
    \includegraphics[width=\textwidth]{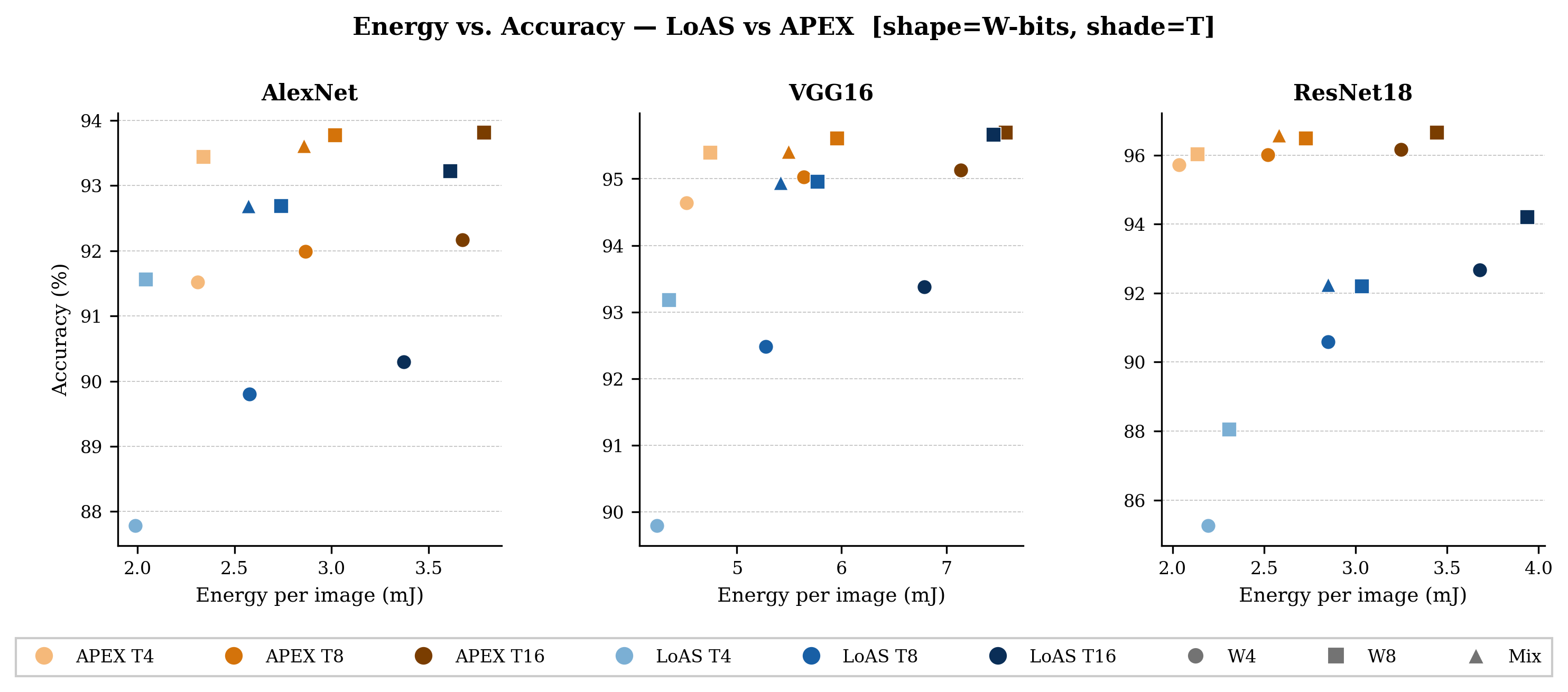}
    \caption{Accuracy vs.\ energy per image for LoAS and APEX accelerators. APEX consistently achieves higher accuracy than LoAS at comparable energy across all SNN models, weight precisions ($W \in \{4,8,\text{Mix}\}$), and timestep configurations ($T \in \{4,8,16\}$), shifting the Pareto frontier of the energy--accuracy trade-off.}
    \label{fig:scatter}
\end{figure*}

Fig.~\ref{fig:energy} compares the inference energy per image of APEX and LoAS for all models and configurations. For ResNet18, APEX achieves lower energy than LoAS for several configurations despite providing higher accuracy. For instance, APEX using W8,T8 achieves 96.49\% accuracy at 2.728\,mJ, compared to LoAS with 92.2\% accuracy at 3.034\,mJ for the same configuration. This energy reduction stems from the higher weight sparsity introduced by the PASC-IF neuron relative to the standard IF neuron under the same quantization setting. For AlexNet and VGG16, APEX incurs a modest energy overhead of up to 16.2\% for the same configuration as LoAS. However, this overhead is offset by substantial accuracy gains: for AlexNet, APEX at W8,T4 achieves 93.4\% compared to 91.56\% for LoAS; for VGG16, APEX at W8,T4 achieves 95.39\% compared to 93.18\% for LoAS --- improvements of 1.88\% and 2.21\%, respectively, at the same energy budget. 

The energy--accuracy trade-off is captured in Fig.~\ref{fig:scatter}, which plots accuracy against energy per image across all models, precisions, and timestep configurations. APEX points consistently appear in the upper-left region compared to their LoAS counterparts, achieving a higher accuracy at equal or reduced energy. This shift is most pronounced for ResNet18, where APEX at W4,T4 achieves 95.72\% accuracy at 2.034\,mJ while LoAS at W8,T8 achieves only 92.20\% at 3.034\,mJ. Across all three models, APEX at W4,T8 matches or exceeds the accuracy of LoAS at W8,T16 while consuming significantly less energy. On average, the best accuracy configurations of APEX achieve 40\% energy reduction compared to LoAS. Mixed-precision configurations further improve this trade-off, with APEX Mix,T8 achieving accuracy comparable to W8,T8 at reduced energy across all models.

For ResNet-34 SNN on ImageNet with INT8 quantization, APEX employing PASC-IF neurons achieves 74.3\%  accuracy at $T{=}8$, whereas LoAS with standard IF neurons incurs an accuracy degredation of approximately 5\% even at $T{=}32$. LoAS consumes 181.76\,mJ per inference, since it requires four times more timesteps, compared to 69.23\,mJ for APEX---a 61.9\% reduction in energy consumption. To match APEX accuracy, LoAS would require $T{=}1024$ timesteps; synthesizing LoAS in this configuration yields a dynamic power consumption of 68.94\,mW, which is on average \textbf{19$\times$ higher} than APEX.

\section{Conclusion}

We presented APEX, a dual-sparse temporal-parallel SNN inference accelerator that integrates the PASC-IF neuron and supports mixed-precision inference. By exploiting the fully temporal-parallel dataflow, all three PASC stages are implemented as combinational circuits completing in 3 cycles, with negligible latency overhead. Across AlexNet, VGG16, and ResNet18 on CIFAR-10, APEX achieves on average 40\% energy reduction over LoAS for best accuracy configurations. For ImageNet, APEX achieves 74.30\% accuracy on ResNet34 SNN at $T{=}8$ while consuming 69.23\,mJ per inference. LoAS, on the contrary, provides ${\sim}$5\% lower accuracy even at $T{=}32$ while incurring a higher energy consumption of 181.76\,mJ per inference. Matching APEX accuracy with QCFS would require $T{=}1024$ timesteps, increasing dynamic power by 19$\times$. These results demonstrate that ANN-equivalent SNN inference is achievable at minimal hardware cost, establishing APEX as an efficient accelerator for precise SNN deployment targeting edge applications.

\bibliographystyle{IEEEtran}
\bibliography{references}

\end{document}